\documentclass[trackchanges]{aastex701}
\usepackage{float}
\usepackage{amsmath}
\usepackage{xcolor}

\begin{document}

\title{Hunting Hidden Axion Signals in Pulsar Dispersion Measurements with Machine Learning}

\correspondingauthor{Guoliang L\"u, Xuefei Chen}
\email{guolianglv@xao.ac.cn, cxf@ynao.ac.cn}


\author[0009-0007-9418-2632]{Haihao Shi}\thanks{Haihao Shi and Zhenyang Huang contribute equally to this work}
 \email{shihaihao@xao.ac.cn}
\affiliation{Xinjiang Astronomical Observatory, Chinese Academy of Sciences, Urumqi 830011, People’s Republic of China}

\affiliation{College of Astronomy and Space Science, University of Chinese Academy of Sciences, Beijing 101408, People’s Republic of China}

\author[0009-0009-8188-5632]{Zhenyang Huang}\thanks{Haihao Shi and Zhenyang Huang contribute equally to this work}%
 \email{huangzhenyang@xao.ac.cn}
\affiliation{Xinjiang Astronomical Observatory, Chinese Academy of Sciences, Urumqi 830011, People’s Republic of China}
\affiliation{College of Astronomy and Space Science, University of Chinese Academy of Sciences, Beijing 101408, People’s Republic of China}

\author[0009-0001-0529-1172]{Qiyu Yan}%
 \email{yanqiyu@st.gxu.edu.cn}
\affiliation{Guangxi Key Laboratory for Relativistic Astrophysics, School of Physical Science and Technology, Guangxi University, Nanning
530004, People’s Republic of China}

\author[0009-0006-9175-5709]{Jun Li}%
 \email{lij@pmo.ac.cn}
\affiliation{School of Physical Science and Technology, Xinjiang University, Urumqi 830046, People’s Republic of China}

\author[0000-0002-3839-4864]{Guoliang L\"u}
\email{guolianglv@xao.ac.cn}
\affiliation{School of Physical Science and Technology, Xinjiang University, Urumqi 830046, People’s Republic of China}
\affiliation{Xinjiang Astronomical Observatory, Chinese Academy of Sciences, Urumqi 830011, People’s Republic of China}

\author[0000-0001-5284-8001]{Xuefei Chen}%
\email{cxf@ynao.ac.cn}
\affiliation{International Centre of SupernovaeInternational Centre of Supernovae (ICESUN), Yunnan Key Laboratory of Supernova Research, Yunnan Observatories, Chinese Academy of Sciences, Kunming 650216, People’s Republic of China}

\affiliation{Key Laboratory for Structure and Evolution of Celestial Objects, Chinese Academy of Sciences, Kunming 650216, People’s Republic of China}


\affiliation{College of Astronomy and Space Science, University of Chinese Academy of Sciences, Beijing 101408, People’s Republic of China}

\date{\today}

\begin{abstract}

In axion models, interactions between axions and electromagnetic waves induce frequency-dependent time delays determined by the axion mass and decay constant. These small delays are difficult to detect, limiting the effectiveness of traditional methods. We compute such delays under realistic radio telescope conditions and identify a prominent dispersive feature near half the axion mass, which appears non-divergent within the limits of observational resolution. Based on this, we develop a machine learning method that achieves \textcolor{black}{90\%} classification accuracy and demonstrates well performance in low signal-to-noise regimes. 
The method's robustness is confirmed against false positives using both simulated noisy data and real-world, known-null observations. 
Future improvements in optical clock precision and telescope bandwidth, particularly with instruments such as the Qitai Radio Telescope, may enhance constraints on the axion decay constant by up to four orders of magnitude in the $10^{-6} \sim 10^{-4}$ eV mass range.
\end{abstract}



\section{Introduction} \label{sec:intro}

Dark matter constitutes approximately 85\% of the Universe’s total matter content \citep{Planck:2018vyg}, as evidenced by a multitude of astrophysical and cosmological observations \citep{Bertone:2004pz,Young:2016ala,Billard:2021uyg}. Despite its pervasive presence, the fundamental nature and composition of dark matter particles remain elusive, pointing to new physics beyond the Standard Model (SM) and general relativity. There are many candidates for dark matter, such as axions \citep{PRESKILL1983127,Abbott:1982af,1983PhLB..120..137D}, fuzzy dark matter \citep{Hu:2000ke,Hui:2021tkt,Burkert:2020laq}, primordial black holes \citep{Lacki_2010,Carr:1974nx,Carr:2016drx}, and so on. Among them,the axion has been a well-motivated DM candidate for a long time \citep{Weinberg:1977ma,Svrcek:2006yi,Hui:2021tkt,Adams:2022pbo}, which were originally proposed to solve the strong CP problem in quantum chromodynamics (QCD) through the Peccei-Quinn mechanism \citep{Peccei:1977hh,Weinberg:1977ma,Wilczek:1977pj,Peccei2008,PhysRevLett.133.161001}. The discovery and confirmation of the Higgs boson’s existence have reinvigorated interest in bosonic dark matter candidates, including axion-like particles (ALPs) predicted by string theory’s rich landscape of scalar fields \citep{Svrcek:2006yi}, often referred to as the “axiverse” \citep{2010PhRvD..81l3530A}. Axions can be produced in the early Universe through various mechanisms \citep{Ringwald:2012hr}, leading to high phase space densities and the potential formation of Bose-Einstein condensates (BECs) \citep{Sikivie:2009qn}. Consequently, axions may aggregate into gravitationally bound structures known as axion stars, which could constitute a significant fraction of dark matter \citep{Braaten:2019knj,Di:2023xaw,Di:2024tlz}. Given the current understanding of galaxy formation, dark matter is considered cold, or equivalently, non-relativistic, massive, and stable—characteristics that align with those of axions. Therefore, axions are strong candidates for what is referred to as “axion dark matter”.

Since galaxies, galaxy clusters, and other cosmic structures typically possess dark matter halos \citep{Navarro:1995iw,Burkert:1995yz,Yang:2011bi,Klypin:2014kpa}, the observational signatures arising from the interaction between dark matter and astrophysical objects are of particular interest. 
Given that pulsars exhibit both high-intensity emission and timing stability comparable to atomic clocks \citep{Lyne_Graham-Smith_2012,NANOGrav:2017wvv,Kaspi:1994hp,10.1093/mnras/stz3071,Becker:2017yyc}, their signals have been extensively studied to probe effects arising from interactions with the dark matter background, such as modifications of X-ray pulsar pulse profiles due to dark matter halos \citep{Liu:2024swd}, gravitational potentials affecting radio pulse travel times \citep{2023PhRvL.131q1001S,losecco2023}, and axion-photon mixing effects \citep{2023PhRvL.131k1004N,Noordhuis:2023wid}. In an astrophysical environment rich in axions and ordinary cosmic media, electromagnetic signals can experience frequency-dependent time delays, manifesting as differences in the arrival times of signals with different frequencies within the same electromagnetic wave. These delays are analogous to the dispersion measure (DM) caused by plasma but include additional contributions from the axion medium \citep{Brevik:2020cky}. The axion field's properties and its interactions with pulsar signals play a crucial role in shaping these delays. By studying these effects, researchers can explore a novel method to constrain the mass, abundance, and interaction strength of axion dark matter, offering valuable insights into the elusive nature of dark matter.

However, the time delays caused by axion-rich environments in pulsar signals are often relatively weak. These signals may be buried in observational noise and difficult to extract due to the low signal-to-noise ratio (SNR). Fortunately, advances in machine learning and neural networks have led to their growing use in astrophysical low-SNR data processing.
In the challenging domain of gravitational wave data analysis, where signals often have low signal-to-noise ratios, trained deep neural networks have demonstrated the ability to extract continuous 30-day inspiral-phase signals from real LIGO data with SNRs as low as 10, and these signals were subsequently used to predict merger times accurate to within 24 hours \citep{wanghe}.
Similarly, in processing \textcolor{black}{radio} telescope data, EMSCA-UNet architectures, which combine convolutional operations with attention mechanisms, have been used to identify radio frequency interference (RFI) in observations of pulsars, demonstrating superior RFI recognition performance \citep{2024MNRAS.529.4719G}.
Moreover, machine learning techniques have been employed to denoise radio astronomical images—effectively removing noise from complex fields to identify and extract faint objects at the limits of instrumental sensitivity \citep{10.1093/mnras/stab3044}, and to denoise spectral line cubes from the IRAM 30 m ORION-B survey, enhancing signal‐to‐noise in low‐signal voxels while preserving spectral shapes \citep{2023A&A...677A.158E}.
These demonstrates the capacity of machine-learning frameworks to learn the distinctive features of radio signals and to distinguish true signal from noise.
Therefore, we can leverage machine-learning frameworks and techniques to develop models capable of Hunting the subtle time delays induced by axion-rich environments in pulsar signals.


In \autoref{Time Delay}, we derive the theoretical framework for calculating the dispersion-induced time delays caused by axion dark matter. 
\autoref{Machine Learning} details the simulation methodologies and the machine learning frameworks employed to model and analyze the pulsar signal data within the axion context. 
In \autoref{Parameter Constraints}, we present the axion parameter space that can be probed. 
\autoref{Discussion} discusses the limitations of our model, evaluates its robustness, and addresses validation against false positives. 
Finally, \autoref{Conclusion} summarizes our results and explores their implications for future astrophysical research and astronomical observations.

\section{Time Delay of Electromagnetic Waves in the Axion Medium}\label{Time Delay}

The time delay function of an electromagnetic wave in a plasma medium is given by \citep{2004hpa..book.....L01}:

\begin{equation}
t(\nu) = \int_{0}^{D_{z}} \frac{dl}{v_g(\nu)} \approx \int_{0}^{D_{z}} \frac{dl}{c} \left( 1 + \frac{1}{2} \frac{\omega_p^2}{\omega^2} \right),
\end{equation}
where $c$ is the speed of light, $\omega_p$ is the plasma frequency, $\omega =2\pi\nu$ is the angular frequency of the electromagnetic wave, and $D_{z}$ is the comoving distance from the source to the detector. The time delay between electromagnetic waves of different frequencies reaching the detector is given by:

\begin{equation}
\Delta t = t(\nu_2) - t(\nu_1).
\end{equation}

We consider a dark matter model composed of axions. Initially proposed to resolve the strong CP problem in QCD, axions have become promising dark matter candidates due to their weak interactions and potential Bose-Einstein condensation in astrophysical environments. However, as electrically charged axions contradict experimental findings, the conventional plasma frequency formula, dependent on free electron density, is inapplicable in this context. Despite this, significant progress in axion electrodynamics \citep{Wilczek:1987mv,2013,Martin-Ruiz:2015skg,Fukushima:2019sjn} has established a strong theoretical foundation for describing axion-photon interactions. This framework is further supported by connections to topological insulators, where axion-like couplings arise naturally in condensed matter systems. Experimental efforts, such as those conducted by the European Organization for Nuclear Research Axion Solar Telescope (CAST), have reinforced the plausibility of axion-induced electromagnetic effects, motivating further exploration of axionic plasmas.

By considering an axionic plasma coupled to photons, \citep{Brevik:2020cky} provide a robust theoretical approach to studying how axion dark matter influences the propagation of electromagnetic waves in astrophysical plasma environments. This perspective opens promising avenues for detecting axion dark matter through astrophysical observations, such as pulsar timing, cosmic microwave background distortions, and radio signals from magnetized astrophysical objects. Ultimately, this model not only enhances our understanding of axion physics but also contributes to the broader effort of uncovering the nature of dark matter through its subtle but potentially detectable effects on cosmic signals.

To obtain the axion corresponding to $\omega_p$, we need to calculate the Feynman diagram for the following process:

\begin{figure}[H]
    \centering
    \includegraphics[width=0.75\textwidth]{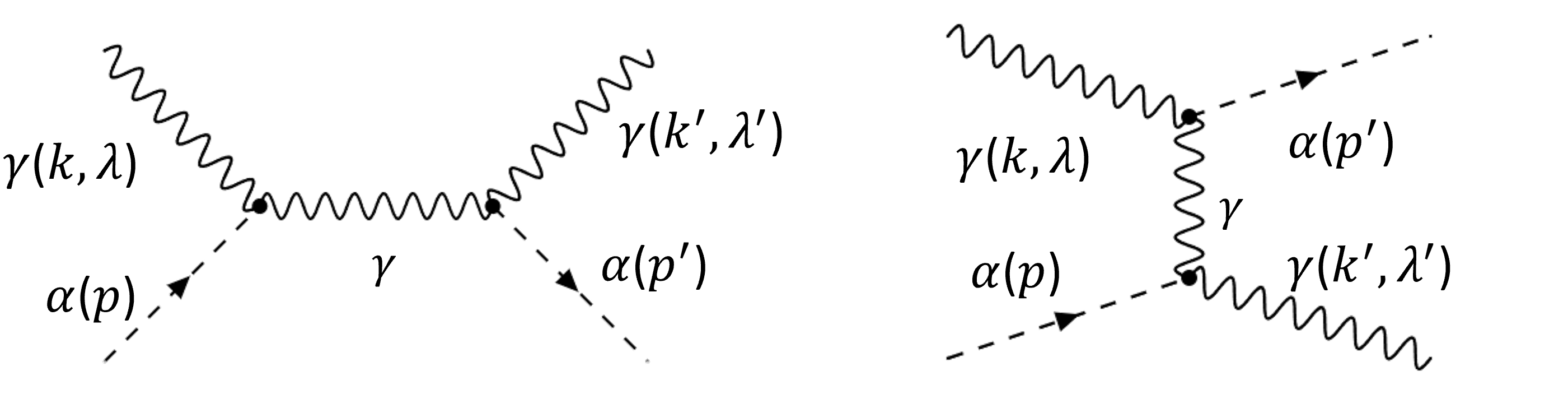}
    \caption{Feynman diagrams for photon scattering on axion. The left panel represents the $s-channel$, where the four-momenta of the incoming photon and axion are $k$ and $p$, respectively. The resulting virtual photon propagates with momentum $k + p$. The outgoing momenta are $k'$ (photon) and $p'$ (axion), satisfying four-momentum conservation: $k + p = k' + p'$. The right panel represents the $u-channel$, which is similar to the $s$-channel, except that the four-momentum of the virtual photon, $k + p$ is replaced by $k - p'$.}
    \label{figFeynman}
\end{figure}

The cross-section is represented as:

\begin{equation}
d \sigma = \frac{k_0^{\prime}}{64 \pi^2 k_0 p_0 p_0^{\prime}} \left(\frac{1}{2} \sum_{\lambda, \lambda^{\prime}} |\mathcal{M}|^2 \right) d \Omega.
\end{equation}

The matrix elements \(\mathcal{M}\) can be obtained from the results in \citep{Brevik:2020cky}. The scattering amplitude for $\theta = 0$ is:

\begin{equation}
\begin{aligned}
\mathcal{M}(0) & = \mathcal{M}_{(s)}(0) + \mathcal{M}_{(u)}(0) \\
& = \frac{3 g^2}{2} \delta_{\lambda^{\prime} \lambda} \left( \frac{k_0^2 m_a}{(2 k_0 + m_a)} + \frac{k_0^2 m_a}{(-2 k_0 + m_a)} \right).
\end{aligned}
\end{equation}

The coupling constant in the Lagrangian term is given by \( g = \frac{\alpha K}{8\pi f_a} \) to align with the convention used in our previous work \citep{PhysRevD.111.023011}, where $\alpha$ is the fine structure constant, $K$ is a model-dependent constant generally expected to be of order 1, and we set $K = 1$ in this article. The parameter $f_{a}$ is the axion decay constant. The mass of the axion is denoted as $m_{a}$.

Thus, the differential cross-section becomes:

\begin{equation}\label{cs}
\begin{aligned}
d \sigma(0) & = \left(\frac{3 g^2}{8 \pi}\right)^2 \left| \frac{k_0^2 m_a}{-4 k_0^2 + m_a^2} \right|^2 d \Omega.
\end{aligned}
\end{equation}

Since the forward scattering amplitude is:

\begin{equation}
|f(0)| = \left( \frac{d \sigma(0)}{d \Omega} \right)^{\frac{1}{2}}.
\end{equation}

The amplitude diverges at \( k_0 = \frac{1}{2} m_a \), which usually requires regularization. However, in an astrophysical context, we interpret this as an indication of a resonance effect. The divergence merely reflects that the quantum field theory approach remains an effective theory, and in a more fundamental framework, this divergence would be resolved. Nevertheless, the resonance effect itself is a real physical phenomenon. In actual observations, radio telescopes do not record continuous frequencies. If the frequency resolution is sufficiently high, the corresponding cutoff may become observable.

Since the photon in the scattering process is free, we obtain:

\begin{equation}
|f(0)| = \frac{3 g^2}{8 \pi} \frac{\omega^2 m_a}{|4 \omega^2 - m_a^2|}.
\end{equation}

Using the relationship:

\begin{equation}
\sqrt{1 - \frac{\omega_p^2}{\omega^2}} = 1 + \frac{2\pi N \text{Re} f(0)}{\omega^2} .
\end{equation}

We consider the radio wavelength band, satisfying \(\omega^2 \gg \omega_p^2\), and the plasma frequency in the axion medium can still be expressed as \(\omega_p^2 = -4\pi N \text{Re} f(0)\) as given by \citep{Brevik:2020cky}, leading to:

\begin{equation}
\omega_p^2 = \frac{3 \rho_{a} g^2}{2} \frac{\omega^2}{|4 \omega^2 - m_a^2|},
\end{equation}
here, $\rho_{a} = N \times m_{a}$ is the mass density of the axion halo, where $N$ is the number density of axions. For radio frequency signals from pulsars, within the region of interest, if the signal energy and axion mass satisfy the relation $\omega = \frac{1}{2}m_a$, a resonance-like effect occurs, resulting in a pronounced peak in the axion plasma frequency at the corresponding signal frequency.


It is important to emphasize that the frequency resolution used in our calculations is 100 kHz, which guarantees that the condition $\omega \gg \omega_p$ holds even as $\omega$ approaches $m_a/2$. This is because our results correspond to an average over the interval $\omega \pm \pi \Delta f$. Furthermore, it can be shown that the condition $\omega \gg \omega_p$ would be violated only when
\begin{equation}
\omega-\frac{m_a}{2}<\frac{7.5 \times 10^{-36} \mathrm{eV}^2}{4 m_a},
\end{equation}
however, due to our adopted frequency resolution and the parameter space considered in this study, our analysis does not enter this regime.

The time delay function obtained in the previous section with the inclusion of \(\omega_p\) is given by:

\begin{equation}\label{dt}
\textcolor{black}{
\Delta t = t(\nu_2) - t(\nu_1) =\int_{0}^{D_z} \left(\frac{\omega_{p2}^2}{2\omega_2^2} - \frac{\omega_{p1}^2}{2\omega_1^2}\right)  \frac{1}{c} \, dl.}
\end{equation}

\textcolor{black}{Without considering cosmological redshift, in \autoref{dt} the only distance-dependent variable in the integral is the density. In addition to smooth transition dark matter profiles such as NFW \citep{1997ApJ...490..493N} and Einasto model \citep{2022A&A...667A..47B}, dark matter may also form clumps along the path from the source to the detector \citep{PhysRevD.103.103514,PhysRevD.100.023506,PhysRevD.82.063531}. However, it is important to clarify the concern about uneven density variations affecting distortions of the resonant feature. In our signal model, which is a single mass axion field, such morphological distortions do not occur. This is because the position of the resonant enhancement is entirely determined by the axion mass ($\omega = m_a/2$). Variations of $\rho_a$ along the path only modulate the magnitude of the time delay but do not shift the resonance frequency. Likewise, under this assumption the intrinsic width of the resonance is determined by the axion–photon interaction cross section (see \autoref{cs}) and by instrumental resolution, not by inhomogeneities in the density along the path. Therefore, the clumpiness of $\rho_a$ affects the strength of the time delay but does not lead to a broadening in frequency space.
}

With a frequency resolution of 100 kHz in our calculation, we observe that when the signal frequency satisfies \( \omega =\frac{1}{2} m_a \), the peak time delay is nearly 9 orders of magnitude larger than its minimum value. Moreover, this contrast increases with higher frequency resolution. Due to the presence of this peak, we can specifically search for characteristic structures in the data, thereby enhancing our sensitivity to time delays. It is noticeable that the time delay equations caused by axions and electrons are different.

\begin{figure}[H]
    \centering
    \includegraphics[width=0.65\textwidth]{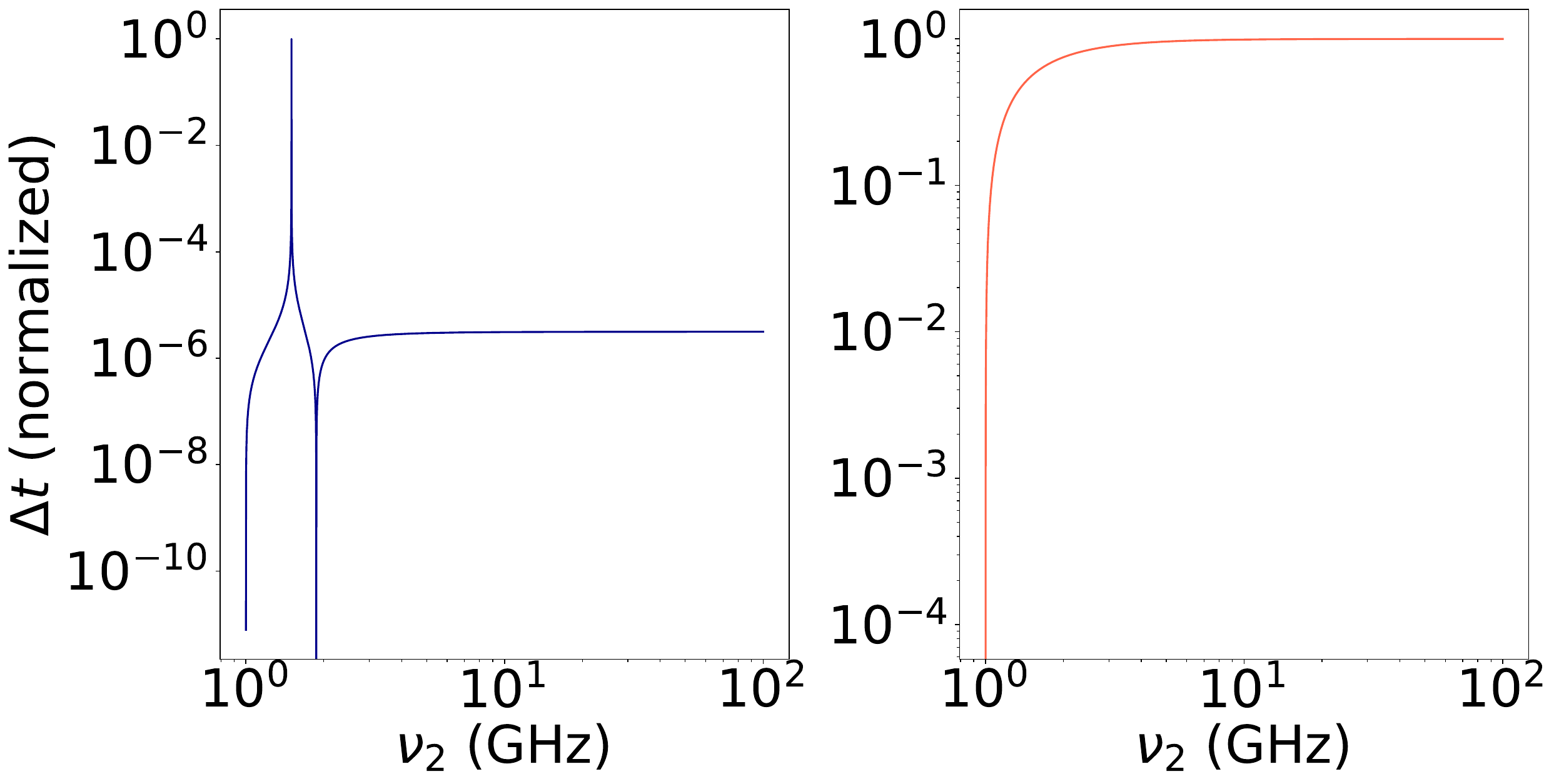}
    \caption{Time delay caused by axion(Left) and electron media(Right) as a function of the incident electromagnetic wave energy. In this calculation,We set \(\nu_1 = 1\) GHz and take \(\nu_2\) as a variable, with a maximum value of 100 GHz. To facilitate a clearer comparison, all time delay curves have been normalized. Here, we set the example axion mass with $m^{'}_{a}$ corresponding to 3 GHz. It can be seen that a prominent peak appears at \(\nu_2 = \frac{1.5}{2\pi}\) GHz, which corresponds to $\frac{1}{2} m^{'}_{a}$.
}
    \label{peak}
\end{figure}

The left panel of \autoref{peak} shows the axion-induced time delay, which differs significantly from the electron-induced delay shown in the right panel. Notably, the axion-induced delay exhibits a prominent frequency-dependent peak, enabling effective identification via machine learning methods discussed in subsequent sections. In other words, our analysis essentially involves searching pulsar dispersion signals for a characteristic frequency exhibiting prominent dispersion, as illustrated in the left panel of \autoref{peak}.

\section{The Machine Learning Method}\label{Machine Learning}  



In light of the characteristic frequency-dependent signatures shown in \autoref{peak}, we aim to determine whether such features can be reliably identified in pulsar dispersion data. However, since no such anomalous delay peaks have been observed in current datasets, we infer that any axion-induced delays—if present—are likely too subtle to be resolved with existing temporal precision. To assess the detectability of such signals under realistic conditions, we construct a machine learning method trained and evaluated on simulated data. This approach, widely adopted in astronomical data analysis, allows us to develop and validate detection strategies in preparation for future high-resolution observations.

\begin{figure}[H]
  \centering
  \begin{minipage}[b]{1\textwidth}
    \centering
    \includegraphics[width=\linewidth]{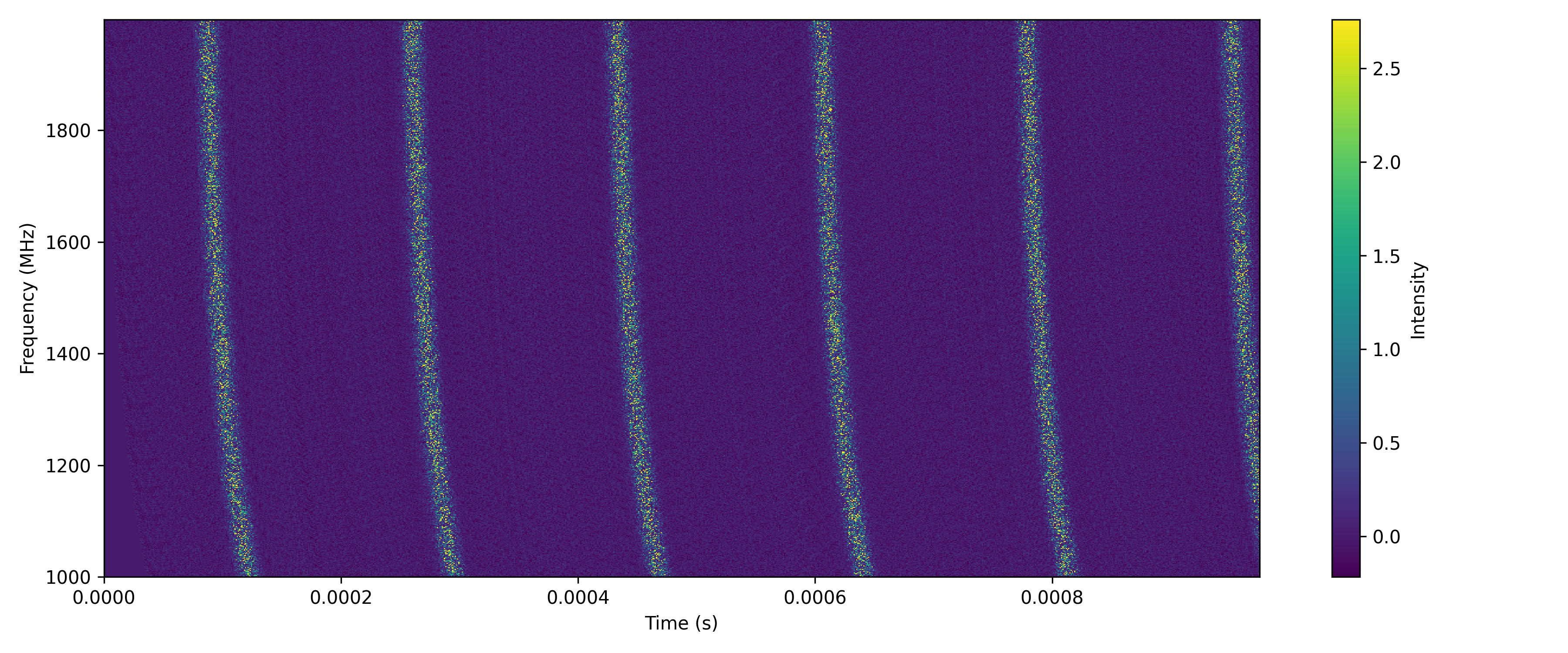}
  \end{minipage}\hfill
  \begin{minipage}[b]{1\textwidth}
    \centering
    \includegraphics[width=\linewidth]{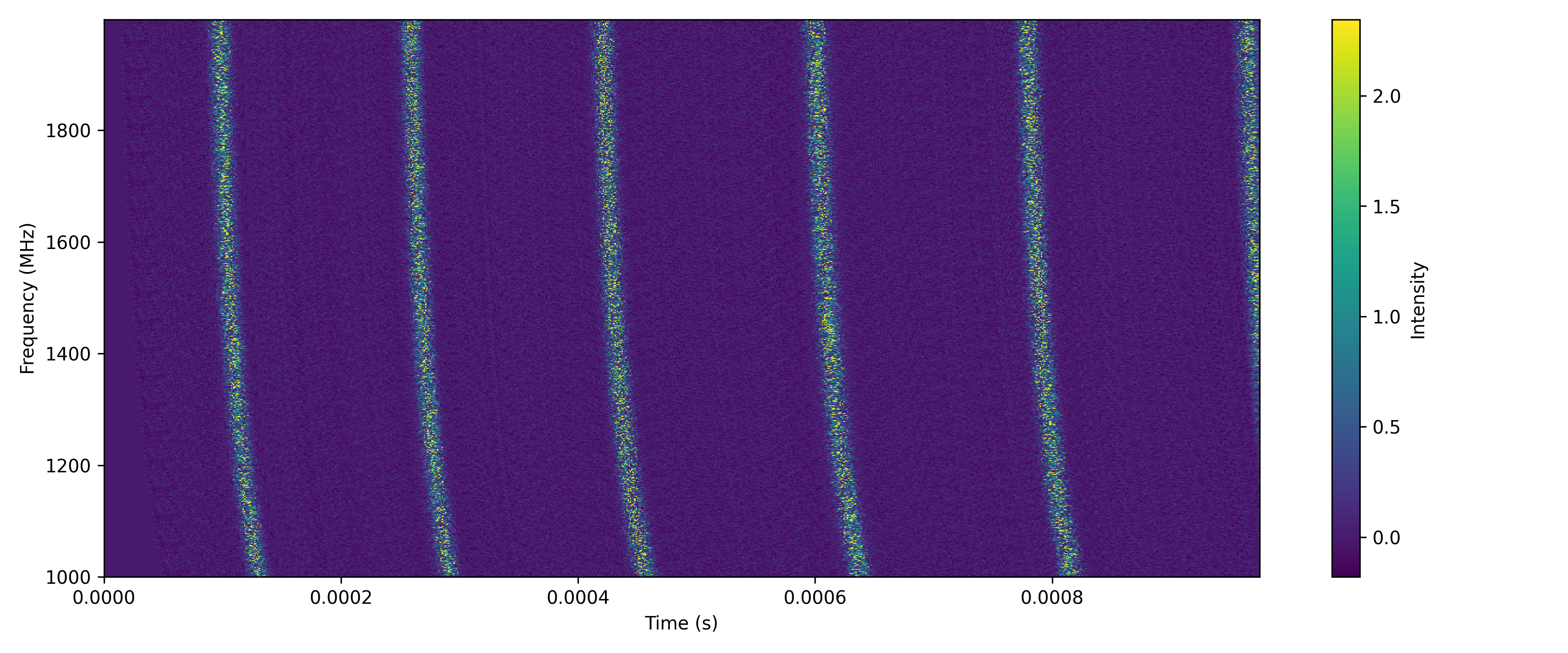}
  \end{minipage}
    \caption{\textcolor{black}{Waterfall dynamic spectra (1–2\,GHz) used to derive time–delay curves.
    Top: baseline simulation including radiometer (thermal white) noise only.
    Bottom: same setup with intrinsic achromatic red timing noise added, modeled as a power–law Fourier–GP and applied as a common time warp.
    Source parameters follow PSR J1933$-$6211 ($P=3.543$\,ms, ${\rm DM}=11.520~{\rm pc\,cm^{-3}}$).
    The band is assembled from 50\,MHz tiles (32 channels each; $f_s=1$\,MHz).
    Intensity is in arbitrary units; both panels share the same color scale.}}
  \label{waterfall}
\end{figure}


\textcolor{black}{To achieve this, we utilized the pulsar signal simulator \texttt{PsrSigSim} \citep{PsrSigSim}. to generate observations for frequencies ranging from 1 GHz to 2 GHz for our experiments.  The package simulates radio telescope observations and supports realistic telescope configurations and measurement uncertainties. 
Our simulations include radiometer (thermal white) noise computed from the GBT parameters via the radiometer equation. We also add an intrinsic achromatic red timing noise component that represents spin noise, so that the model captures the dominant stochastic contributions beyond the instrument model. \citep{Shannon_2010,10.1093/mnras/sts486}
As a representative target, we chose PSR J1933$-$6211 with $P=3.543~\mathrm{ms}$ and $\mathrm{DM}=11.520~\mathrm{pc\,cm^{-3}}$ for our simulation.
\autoref{waterfall} presents one representative realization.
After generating simulated data that include intrinsic achromatic red timing noise, we used \texttt{PINT} \citep{PINT} to model the red noise component as a stationary Gaussian process with a power-law power spectral density implemented through a Fourier-sum basis. We estimated the red noise hyperparameters together with the deterministic timing model using generalized least squares, then subtracted the fitted red noise contribution from the residuals. From the cleaned data we constructed the frequency–dependent arrival time curves, aligning the workflow with standard pulsar timing practice and bringing the experiment closer to real-world processing.}

\begin{figure}[H]
  \centering
  \begin{minipage}[b]{0.5\textwidth}
    \centering
    \includegraphics[width=\linewidth]{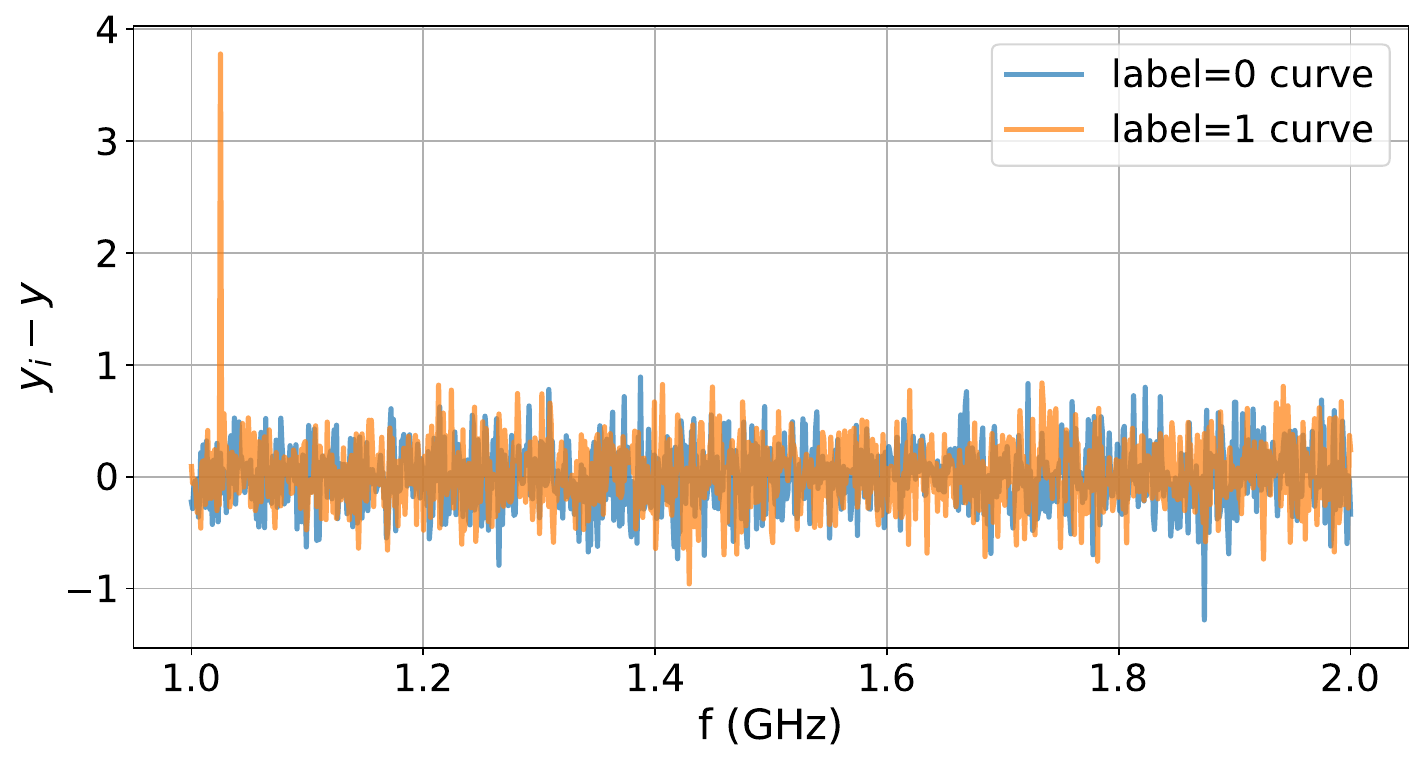}
  \end{minipage}\hfill
  \begin{minipage}[b]{0.5\textwidth}
    \centering
    \includegraphics[width=\linewidth]{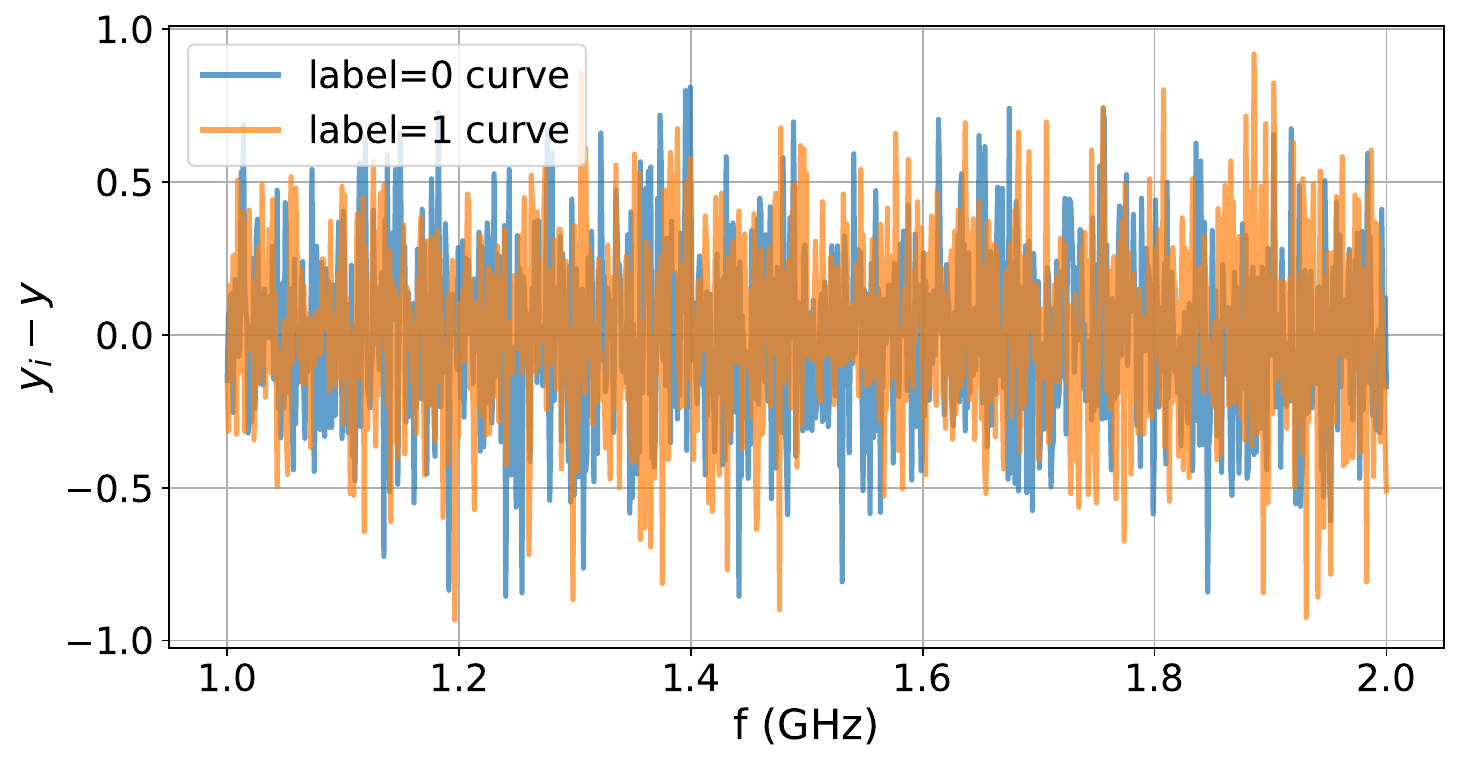}
  \end{minipage}
  \caption{Comparison of time-delay signals under different axion parameters. In left panel, the data labeled “1” represent a time-delay signal induced by axion with parameters \(\bigl(f_{a}^{-1}, m_{a}\bigr)=\bigl(10^{-10}\,\mathrm{GeV}^{-1},1.3\times10^{-6}\,\mathrm{eV}\bigr)\).
  As shown, the SNR for these data points is significantly greater than 1, indicating that the signal can be easily identified using simple peak-search methods. 
  In right panel, the data labeled “1” correspond to a time-delay signal caused by axion with parameters \(\bigl(f_{a}^{-1}, m_{a}\bigr)=\bigl(10^{-10}\,\mathrm{GeV}^{-1},1.7\times10^{-6}\,\mathrm{eV}\bigr)\). 
  Here the SNR lies in the range \(\sim10^{-1}\) to \(\sim10^{0}\), making the signal nearly indistinguishable from white noise and challenging for traditional detection algorithms.
  }
  \label{SNR}
\end{figure}

\textcolor{black}{
We then simulated three types of datasets: (1) a denoise curve without dark matter signal, denoted as \(\{y_0\}\); (2) a denoise curve with a dark matter signal, denoted as \(\{y_1\}\); and (3) the theoretical curve without noise and dark matter signal, denoted as \(\{y\}\). }Since the time-delay effect induced by axions and the pulsar dispersion effect are simply additive, we subtract the theoretical curve \(\{y\}\) from both \(\{y_0\}\) and \(\{y_1\}\). 
This allows for a more intuitive visualization of the impact of white noise on the detection of axion-induced signals. 
\autoref{SNR} illustrate the significance of simulated signals embedded in noise for two sets of axion parameters. Specifically, the data labeled “1”, after subtracting the noise-free theoretical curve, reveal both the observed white noise and the axion-induced time-delay peak, whereas the data labeled “0” yield only the pure observed white noise after the same subtraction.
In particular, our definition of SNR refers to the ratio between the signal peak of the axion-induced delay and the background noise.

\begin{figure}[H]
    \centering
    \includegraphics[width=0.65\textwidth]{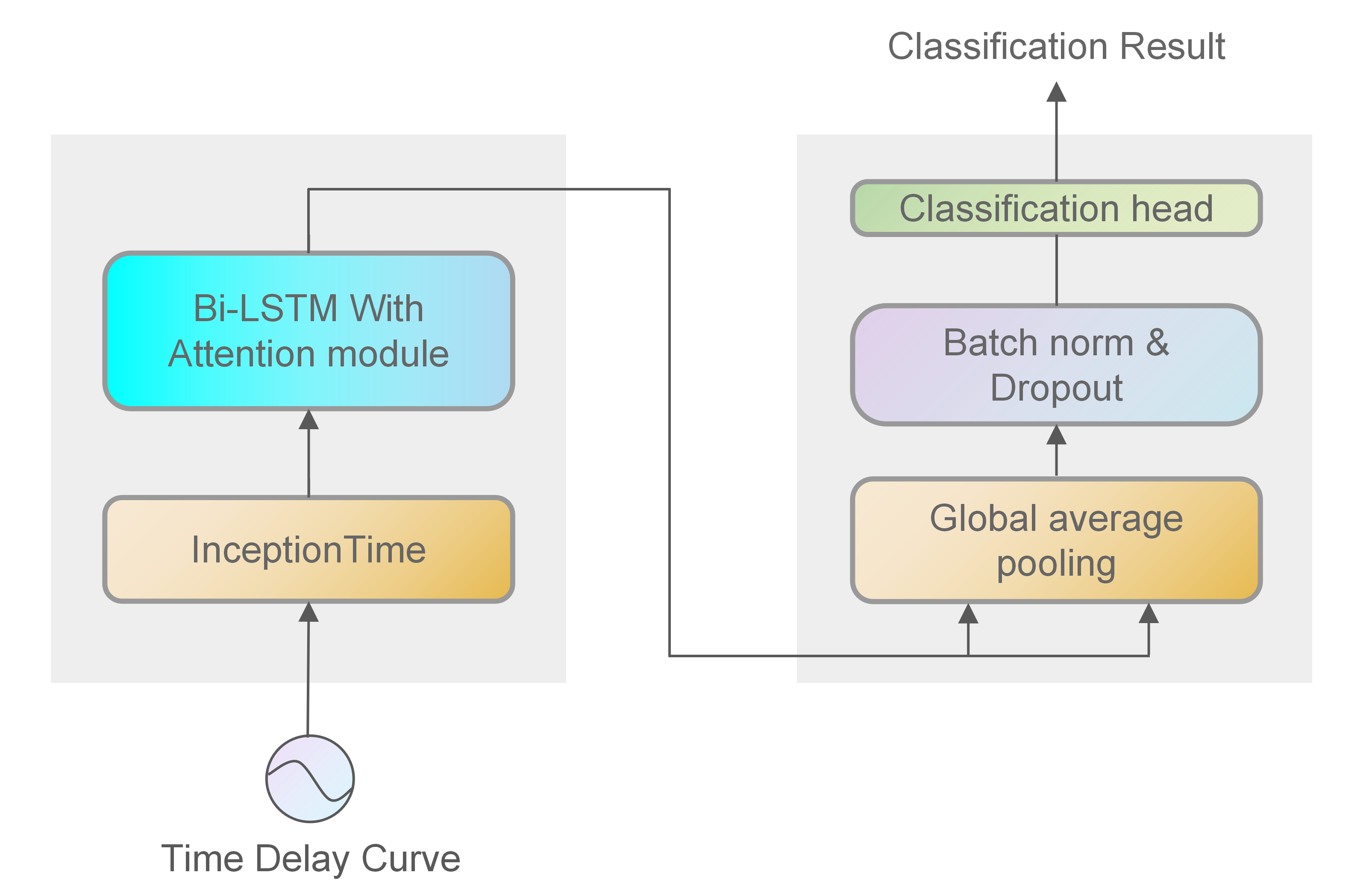} 
    \caption{Schematic diagram of the attention-enhanced InceptionTime network architecture. Building upon the classical multi-scale temporal structure of InceptionTime, the model integrates a temporal receptive field pyramid formed by stacked heterogeneous convolutional kernels and a temporal attention module incorporating BLSTM. These components establish a dynamic weight allocation mechanism to enhance spatiotemporal localization capability for transient peaks. Global average pooling combined with dual regularization layers (batch normalization and dropout) further improves noise robustness. Notably, during the unsupervised pre-training phase, the classification head is replaced by a reconstruction decoder, forming an autoencoder architecture. This stage is designed to learn robust feature representations from the unlabeled data by minimizing a reconstruction loss. Subsequently, these learned features are fed into a clustering algorithm to group the signals into distinct categories.}
    \label{network}
\end{figure}

Fortunately, with the advancements in machine learning, even weak signals—like those shown in the right panel of \autoref{SNR}—that are buried in noise can be identified and extracted by training a suitable machine learning model. 
To this end, we developed an attention-enhanced InceptionTime network, as shown in \autoref{network}, specifically tailored to detect transient peak signals against a white noise background.
Building on the classical multi-scale temporal architecture of InceptionTime \citep{Inception}, our model innovatively incorporates a time-sensitive feature recalibration mechanism. 
This is achieved by stacking multiple groups of heterogeneous convolutional kernels to form a temporal receptive field pyramid that concurrently captures waveform patterns of various durations. 
Subsequently, a temporal attention module \citep{AttentionIsAllYouNeed} is introduced, employing a Bidirectional Long Short-Term Memory (BLSTM) \citep{BLSTM} to encode contextual dependencies across time steps and generating a dynamic weight matrix along the time dimension via sigmoid gating. 
This mechanism enables the network to autonomously concentrate on the sensitive time intervals where spike signals occur, while suppressing noise interference in non-critical regions. 
After temporal attention weighting, a global average pooling compresses the temporal dimension, retaining the overall distribution characteristics of the pulse signal. 
To further enhance noise robustness, batch normalization and dropout regularization (with a dropout rate of 0.5) are integrated during the feature aggregation phase, thereby reducing the risk of overfitting through dual constraints. 
Finally, a fully connected layer is utilized to perform binary classification, discerning between “peak-containing signal” and “pure noise”. 
The synergistic interplay of the temporal attention mechanism and multi-scale feature extraction significantly enhances the model’s spatiotemporal localization capability for transient peak signals under low signal-to-noise conditions, rendering its time-sensitive characteristics particularly suitable for tasks involving the discrimination of singular peak events.

\begin{figure}[H]
    \centering
    \includegraphics[width=0.65\textwidth]{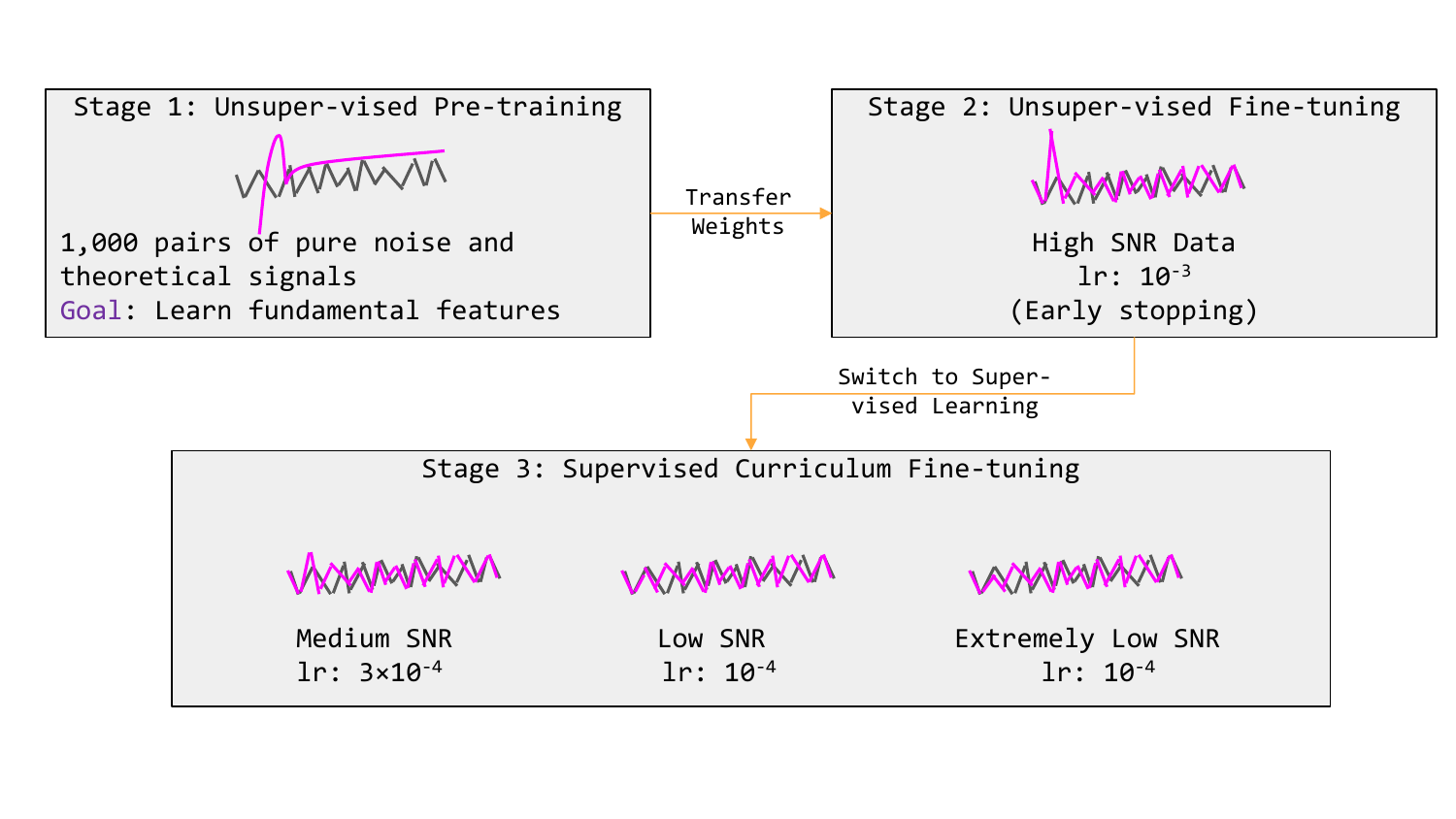} 
    \caption{Schematic diagram of the multi-stage training protocol. The process commences with an unsupervised pre-training stage on theoretical signals and pure noise to learn fundamental features. This is followed by an unsupervised fine-tuning stage on High SNR data (lr: $10^{-3}$) with an early stopping criterion. Finally, a methodological shift to supervised learning is made for the curriculum fine-tuning phase, where the model is sequentially trained on datasets with progressively decreasing signal-to-noise ratios: Medium (lr: $3 \times 10^{-4}$), Low (lr: $10^{-4}$), and Extremely Low (lr: $10^{-4}$).
}
    \label{train_process}
\end{figure}

Our model was trained using a sophisticated, multi-stage curriculum learning strategy designed to progressively enhance its performance from high to extremely low SNR regimes, as shown in \autoref{train_process}.
The training protocol commenced with an unsupervised pre-training stage. We generated a foundational dataset of 1,000 pairs, each containing a pure noise curve and its corresponding noiseless theoretical signal. The model was trained on this dataset to learn the fundamental characteristics of the axion-induced signal and noise. 
This phase concluded once the model's accuracy on the validation set achieved 100\% and remained stable for five consecutive epochs.
Following this pre-training stage, the model was fine-tuned on the ``High SNR'' dataset (detailed in \autoref{table:snr_performance}), continuing with the unsupervised approach. To prevent overfitting and ensure generalization, we employed an early stopping criterion, halting the training after the validation accuracy showed no improvement for five consecutive epochs.
The unsupervised training phase is configured to learn robust feature representations using an autoencoder architecture. In this stage, the model is trained to reconstruct its input by minimizing the Mean Squared Error as the reconstruction loss, using the Adam optimizer with a learning rate of $10^{-3}$. Subsequently, the trained encoder is utilized as a feature extractor, and the resulting feature vectors are clustered via the K-Means algorithm to group the signals into distinct categories.

A methodological shift was introduced for the ``Medium SNR'' dataset. We observed that the unsupervised framework struggled to yield high accuracy in this more challenging regime. Consequently, we transitioned to a supervised learning paradigm. The model was trained for a fixed 50 epochs, and to facilitate stable convergence, the learning rate was annealed from $1 \times 10^{-3}$ to $3 \times 10^{-4}$.
In the final stages, the model's robustness was further honed on the most challenging datasets. The learning rate was reduced to $1 \times 10^{-4}$ to enable fine-grained adjustments. The model was then trained sequentially, first on the ``Low SNR'' dataset and subsequently on the ``Extremely Low SNR'' dataset, with each stage lasting for 50 epochs.
Throughout all training stages, the datasets were consistently partitioned into 80\% for training and 20\% for validation. The volume of data for each SNR category (High, Medium, Low, and Extremely Low) was kept uniform. The AdamW optimizer was used for all supervised stages, with a weight decay of $10^{-4}$, to minimize the cross-entropy loss function.
To evaluate the model's performance, we tested it on subsets of the test set with varying SNR. The detailed performance metrics, including accuracy, F1-score, and recall, are shown in \autoref{table:snr_performance}.
Additionally, detailed training results, procedures, and network architecture can be found in \citep{code}.



\begin{deluxetable}{ccccc}
\tablecaption{Model performance on the test set across different SNR categories. The test set contains an equal proportion of samples from each category.\label{table:snr_performance}}
\tablewidth{0pt}
\tablehead{
\colhead{SNR Category} & \colhead{Accuracy} & \colhead{F1-score} & \colhead{Recall} & \colhead{\textcolor{black}{FPR}}
}
\startdata
High (1.5-2.0) & \textcolor{black}{100\%} & \textcolor{black}{100\%} & \textcolor{black}{100\%} & \textcolor{black}{0.0\%}\\
Medium (1.0--1.5) & \textcolor{black}{92\%} & \textcolor{black}{88\%} & \textcolor{black}{86\%} & \textcolor{black}{4.9\%}\\
Low (0.5--1.0) & \textcolor{black}{85\%} & \textcolor{black}{80\%} & \textcolor{black}{82\%} & \textcolor{black}{13.3\%}\\
Extremely Low (0.1--0.5) & \textcolor{black}{83\%} & \textcolor{black}{81\%} & \textcolor{black}{80\%} & \textcolor{black}{14.5\%}\\
\enddata
\tablecomments{
\textcolor{black}{Definitions: Accuracy $=(TP+TN)/N$; Recall (TPR) $=TP/(TP+FN)$; 
Precision $=TP/(TP+FP)$; F1-score $=2\,(\mathrm{Precision}\cdot\mathrm{Recall})/(\mathrm{Precision}+\mathrm{Recall})$; 
FPR $=FP/(FP+TN)$. }}
\end{deluxetable}

\section{Observation Parameter Constraints}\label{Parameter Constraints}  

We next evaluate the axion parameter space that future radio telescopes can constrain using the method outlined above.
 In the calculations, we choose the following parameters: $\rho_{a} = \rho_{\text{DM}_{\text{local}}} = 0.38\text{GeV/cm}^{3}$ \citep{McMillan:2016jtx}, $\nu_1 = 1\text{GHz} \sim 6.58 \times10^{-7}\text{eV} $, $\nu_2 = 2\text{GHz} \sim 1.316\times10^{-6}\text{eV} $. \textcolor{black}{We use $\rho_{\text{DM}_{\text{local}}}$ as a conservative baseline for our signal generation. The true dark matter distribution in the Milky Way is non-uniform, and for pulsars located towards the Galactic Center, the integrated density along the line of sight is generally expected to exceed this local value. This implies that our simulated signals represent a lower limit of the potential real signal strength. Consequently, the performance of our model is evaluated under a conservative assumption, ensuring its robustness when encountering the more prominent signals expected from actual observations.} For the calculation of comoving distance, since we are focusing on pulsars within the Milky Way, the effects of cosmic expansion are neglected. Setting the redshift to $10^{-6}$ (corresponding to the Hanoi galaxy), the corresponding distance is $2.64\times 10^{20} \text{m}$. The next-generation radio telescopes (e.g., the Square Kilometre Array, SKA) are expected to achieve a timing precision of $10^{-8}$ s for pulsar signal time-of-arrival (TOA) measurements with 10-minute integrations \citep{Liu:2011cka}. Only time delays exceeding this threshold can accumulate, meaning that only the parameter space above this limit can be probed.

The precision and uncertainty of pulse time-of-arrival (TOA) measurements depend on the target source, system parameters, and observation time. The TOA measurement uncertainty ($\sigma_{\text{TOA}}$) is given by \citep{2004hpa..book.....L02}:

\begin{equation}
\sigma_{\text{TOA}} \simeq \frac{W}{SNR} \propto \frac{S_{\text{sys}}}{\sqrt{t_{\text{obs}} \Delta \nu}} \times \frac{P \delta^{3/2}}{S_{\text{mean}}},
\end{equation}
where $W$ is the pulse width, $\mathrm{SNR}$ is the signal-to-noise ratio, and $S_{\text{sys}}$ is the system equivalent flux density. $\Delta \nu$ is the observing bandwidth, $t_{\text{obs}}$ is the integration time, $P$ is the pulsar period, and $\delta = W/P$ is the pulse duty cycle. $S_{\text{mean}}$ denotes the mean flux density of the pulsar.

The TOA uncertainty decreases with increasing SNR, which is given by \citep{2017isra.book.....T}:
\begin{equation}
SNR = C \frac{T_{A}}{T_{S}} \sqrt{ t_{\text{obs}}\Delta \nu},
\end{equation}
where $C$ is a constant greater than or equal to one, $T_{A}$ is the component of the antenna temperature resulting from the target source, and $T_{S}$ is the system temperature. 

It is evident that the precision of pulsar signal observations depends on the observation duration. Existing and planned radio telescope facilities provide the necessary conditions for long-term observations of specific sources in the future. Of course, observational precision also depends on instrumental parameters; however, we do not discuss engineering and technical aspects here.

\begin{figure}[h]
    \centering
    \includegraphics[width=0.6\textwidth]{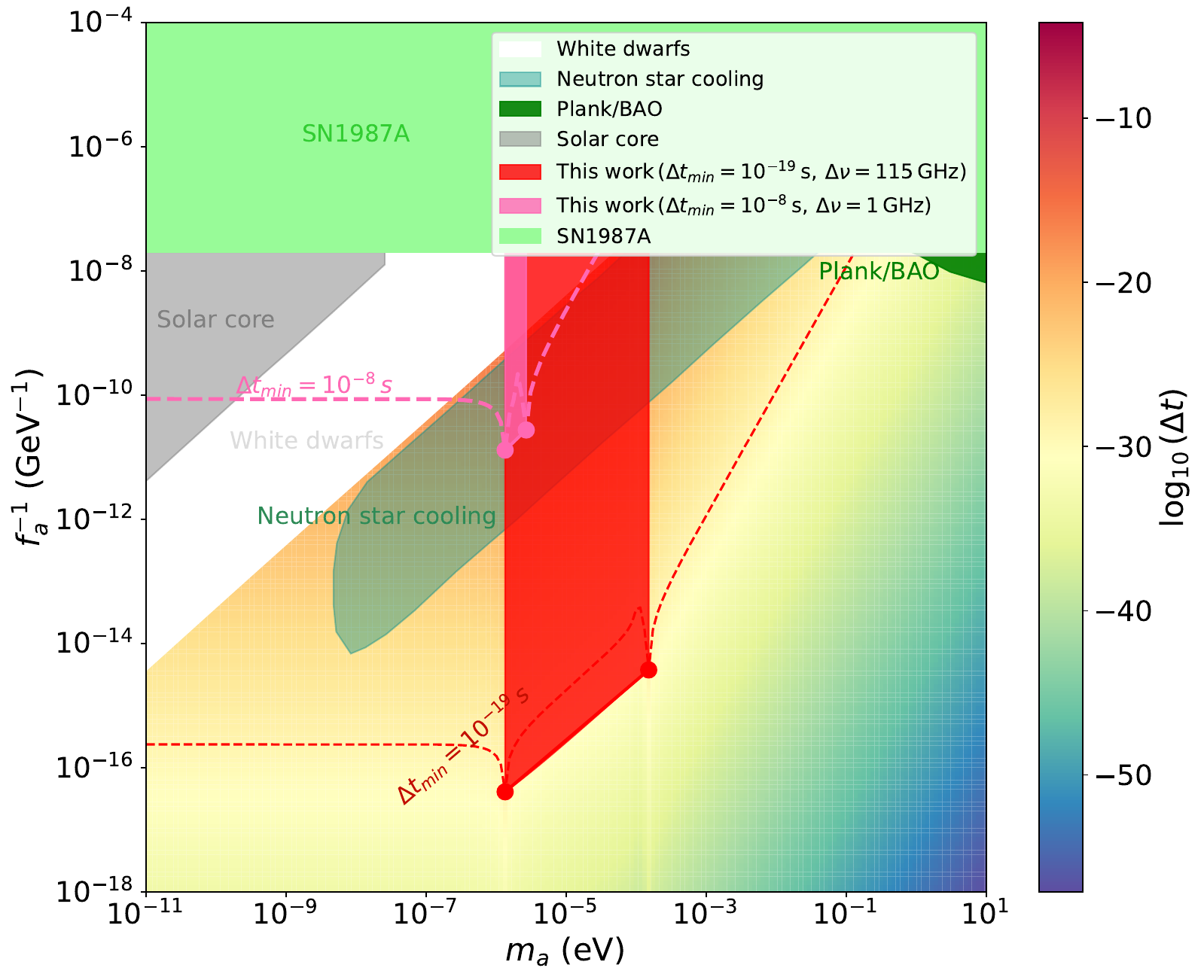}  

    \caption{The pink region in the figure represents the parameter space constrained by the lower precision (with the observation frequency ranging from 1 GHz to 2 GHz). The red region represents the parameter space expected to be scanned in the future (time resolution of \(10^{-19}\) s and bandwidth of 115 GHz). The figure also shows constraints from the solar core (grey) \citep{Hook:2017psm}, white dwarfs (white) \citep{Balkin:2022qer}, neutron star cooling (cyan) \citep{PhysRevLett.133.251002,Kumamoto:2024wjd}, SN1987A (light green) \citep{Springmann:2024ret}, and Planck/BAO (dark green) \citep{Caloni:2022uya}, with data obtained from \href{https://github.com/cajohare/AxionLimits/tree/master}{Axionlimits} project \citep{AxionLimits}. $\Delta t_{\text{min}}$ refers to the minimum time resolution that the device can detect.}  
    \label{1e-8}  
\end{figure}

First, we compute $m_{a}$ within the range of $10^{-11} \text{ eV}$ to $10^{1} \text{ eV}$ and $f_{a}^{-1}$ within $10^{-18} \text{ GeV}^{-1}$ to $10^{-4} \text{ GeV}^{-1}$ to obtain the time delay matrix. This allows us to determine the upper limit and minimum precision of time measurements from existing pulsar data. The parameter space of $m_{a}$ and $f_{a}^{-1}$ above the minimum precision corresponds to regions that can be distinguished or excluded based on current data, as indicated by the pink-shaded region in \autoref{1e-8}, representing the parameter constraints at a time precision of $10^{-8}$ s.

Next, we discuss the precision that may be achievable in the future. Current optical clocks have reached frequencies of \(10^3\) THz \citep{10.1093/nsr/nwaa119}. Over long observation periods, the resolution is determined by the system's uncertainty. The most advanced optical clocks can currently achieve a resolution of 18$\sim$19 significant digits ($10^{-18} \sim 10^{-19} \text{s}$) over observation timescales on the order of days \citep{PhysRevLett.120.103201}. However, this represents the theoretical upper limit under ideal conditions. In practical measurements, additional systematic errors, such as detector delay and cable delay, must also be considered. Here, we only provide the theoretical limit. When applied to future radio telescopes, this would enable the exploration of a much larger parameter space. Although this would significantly increase storage costs, our focus here is solely on the scientific implications.

Regarding bandwidth configuration, we consider future wide-band radio telescopes, such as the Qitai Radio Telescope (QTT), which is expected to achieve a bandwidth of up to 115 GHz \citep{Wang:2023occ}. The parameter space accessible with a 115 GHz bandwidth is shown as the red-shaded region in \autoref{1e-8}.

In \autoref{1e-8}, it can be seen that in the axion mass range of $10^{-6} \sim 10^{-4}\text{eV}$, future high-time-resolution and wide-bandwidth radio telescopes are expected to improve constraints on $f_{a}$ by approximately 4 orders of magnitude compared to current limits.

\section{Discussion}\label{Discussion}

\subsection{limitation}

In this work, we have introduced a novel machine learning method designed to search for transient signals in pulsar DM data, specifically targeting the unique signature of axion-photon resonance. 
Our core finding is that these subtle characteristic resonant peaks can be effectively identified by neural network.
While promising, our current work has several limitations that pave the way for future refinements and investigations.
\textcolor{black}{The main limitation of this study is that the training data do not include red noise introduced by the instrumentation. Our simulations contain radiometer white noise and intrinsic pulsar red timing noise. The intrinsic component has well characterised statistics and can be modelled with \texttt{PINT}, as demonstrated by comprehensive pulsar timing analyses \citep{Arzoumanian2018}. We fit and subtract this component and reduce it to near the noise floor, which leaves residuals dominated by white noise. Under these conditions our network identifies axion signals with good reliability.}

\textcolor{black}{
In contrast to intrinsic timing noise, instrumental and system-dependent low frequency processes arise from instabilities in observatory time standards, errors in the adopted solar system ephemeris, and polarization calibration imperfections as well as other receiver/back-end systematics that vary across telescopes and frequency bands. Clock and ephemeris errors imprint array-level spatial correlations that are difficult to suppress and can masquerade as common low frequency processes \citep{2016MNRAS.455.4339T}. 
For a single radio telescope, slow bandpass variations and low-bit digitization artefacts introduce frequency-dependent profile distortions that bias template matching and degrade TOA precision; in severe cases the resulting 2-bit distortion cannot be corrected post detection. \citep{10.1111/j.1365-2966.2011.19452.x}
Imperfect polarization calibration and time variable cross-coupling in the receiver chain distort the integrated pulse profile and shift the template alignment, thereby introducing systematic biases in the measured TOA that are not fully removed by standard calibration schemes \citep{10.1093/mnras/stv1722}. 
Even with explicit per-system and per-band modeling, residual structure remains and can shift template-aligned TOA, so complete removal is not generally achievable.
We experimented with injecting such disturbances into our simulations, and the outcome was detrimental: the red component could not be removed effectively by our denoising pipeline, and the residuals may obscured even the canonical dispersive delays within modest bandwidths ($<100\,\mathrm{MHz}$). Under these circumstances, distinguishing the presence of an axion signal was essentially infeasible. 
However, recent results show that instrument-aware calibration pipelines can progressively capture system-induced red processes. 
In particular, \cite{Rogers_2024} demonstrates at the 64m Murriyang radio telescope that measurement equation polarimetric calibration combined with matrix template matching improves TOA precision and mitigates instrumental systematics, indicating steady progress toward more complete modelling.
This makes it possible in the future to completely eliminate biases, including instrumental red noise.
Because the principal objective of this work is to articulate a theoretically observable axion electromagnetic resonance signature and to introduce a practical detection strategy, we deliberately simplified instrumental red–noise systematics so that our machine–learning approach can function as a feasible proof of concept.}

\subsection{Model Robustness and Validation Against False Positives}

A critical aspect of any signal-hunting method is its robustness against false positives. To rigorously test and validate our model's reliability, we performed a series of validation tests designed to probe its response to noise and known-null data.

First, to ensure the model does not misinterpret random fluctuations as a signal, we conducted a negative control test. For this, we took a large set of our simulated pulsar signals that were explicitly labeled as containing no axion signal (i.e., the `label 0' training set) and injected an additional layer of strong, random Gaussian noise. This process created a challenging dataset of augmented, `no-signal' cases under unusually noisy conditions. When this dataset was fed into the trained network, the model yielded ambiguous and inconclusive results. This outcome is highly desirable, as it demonstrates that our model is not easily fooled into creating a false positive, even when the noise level is artificially exaggerated. 
Motivated by this robust behavior, for our model, we propose a stringent, ensemble-based confidence criterion for any future detection claim: for a given pulsar, a positive detection can only be asserted if, out of a large ensemble of observations (e.g., $\ge$1000 independent epochs), a supermajority of $>$95\% of the samples are individually classified as containing an axion-induced signal. This criterion ensures that any potential detection is persistent and statistically significant, not the result of a few random, outlier fluctuations.

Next, we moved from simulated noise to the complexities of real astronomical data. It is crucial to emphasize that this real-world dataset does not perfectly match the idealized scenario of our training data. A key simplification in our current work is the exclusion of instrumental red noise from our simulations. However, this type of noise, originating from the telescope system, is an unavoidable component of real observations and cannot be perfectly removed.
This discrepancy makes the test even more significant. We are confronting our model with a dataset containing types of correlated noise it has never been trained on. This serves as a crucial validation step, as the frequency range of this observation corresponds to an axion parameter space that has already been robustly excluded by other experiments, making it a real-world, known-null case. 
Subsequently, we tested our method using observational data of the PSR J1933-6211 pulsar, with frequencies ranging from 982 MHz to 1782 MHz, obtained from CSIRO \citep{J1933}. We trained a model based on the resolution and frequency range of the observation datas to use in our method. 
Our trained model indeed returned a null result, consistently classifying the observations as signal-free.  By correctly identifying a known-null case despite the presence of un-modeled systematic red noise, we strongly demonstrate our method's capability to avoid false positives. It shows the model is robust not only to statistical white noise but also to the intricate, real-world noise complexities that it was not explicitly trained to recognize.

\section{Conclusion}\label{Conclusion}

Following the methodology proposed in \citep{Brevik:2020cky}, we calculate the small time delays induced by the interaction between axions and electromagnetic waves. We find that the time delay undergoes a resonant enhancement when the electromagnetic wave frequency equals half the axion mass. While a strict equality would theoretically lead to a divergence that invalidates the assumptions underlying the calculation, the finite observational bandwidth and time resolution ensure that this resonance remains finite and physically valid within the parameter space accessible to realistic observations. Our results illustrate the axion parameter space that can be constrained by searching for potential additional dispersion effects in radio signals under both current and future precision levels. With the precision of next-generation radio telescopes, constraints on \( f_{a} \) in the axion mass range of \( 10^{-6} \sim 10^{-4} \) eV are expected to improve by approximately 4 orders of magnitude compared to existing methods. To detect these faint signatures, we developed a machine learning-based method capable of identifying axion-induced signals embedded in observational noise. 
\textcolor{black}{We simplified the instrumental red noise and used PsrSigSim to simulate realistic pulsar observations, incorporating instrumental white noise and intrinsic pulsar red timing noise. After processing with PINT-based denoise pipeline, our model reliably determines whether an axion signal is present and recovers subtle features that traditional algorithms struggle to detect.}
We further tested the method on real pulsar data (PSR J1933-6211), finding no evidence of axion-induced time delays within the sensitivity limits.

The effectiveness of our approach is contingent on future improvements in observational precision, particularly through the deployment of higher-accuracy optical clocks and next-generation radio telescopes with wider bandwidths such as QTT. 
Moreover, while our method effectively identifies axion signals concealed within white noise, its performance is dependent on the extent to which red noise has been mitigated in the observational data. 
Future work will focus on further optimizing the method to improve the accuracy of identification, using higher temporal resolution pulsar signal time delay data for axis parameter limiting.

\begin{acknowledgments}

I would like to thank Haoran Di, Kaifan Ji, Xuwei Zhang, and Junda Zhou for the useful discussions during the research process. This work received support from the National Natural Science Foundation of China under grants 12288102, 12373038, 12125303, 12090040/3, and U2031204; the Natural Science Foundation of Xinjiang No. 2022TSYCLJ0006; the science research grants from the China Manned Space Project No. CMS-CSST-2021-A10; the National Key R\&D Program of China No. 2021YFA1600401 and No. 2021YFA1600403; the Natural Science Foundation of Yunnan Province Nos. 202201BC070003 and 202001AW070007; the International Centre of Supernovae, Yunnan Key Laboratory No. 202302AN360001; and the Yunnan Revitalization Talent Support Program $-$ Science \& Technology Champion Project No. 202305AB350003.

\end{acknowledgments}


\bibliography{sample701}{}
\bibliographystyle{aasjournalv7}



\end{document}